\documentclass[jog]{igs}

\usepackage[utf8]{inputenc}

\usepackage{graphicx}
\usepackage{siunitx}
\usepackage{hyperref}
\usepackage{enumitem}
\usepackage{fancyhdr}

\jourvolume{}
\jourissue{}
\jourpubyear{2025}

\begin{document}

\title[]{Glacier data assimilation on an Arctic glacier: Learning from large ensemble twin experiments}
\author[]{Wenxue CAO$^1$,
 Kristoffer AALSTAD$^1$, Louise S. SCHMIDT$^1$, Sebastian WESTERMANN$^1$, and Thomas V. SCHULER$^1$}

\affiliation{%
$^1$Department of Geosciences, University of Oslo, Oslo, Norway\\
  \email{wenxue.cao@geo.uio.no}}

\begin{frontmatter}

\maketitle

\begin{abstract}
Glacier modeling is crucial for quantifying the evolution of cryospheric processes. At the same time, uncertainties hamper process understanding and predictive accuracy. Here, we suggest improving glacier mass balance simulations for the Kongsvegen glacier in Svalbard through the application of Bayesian data assimilation techniques in a set of large ensemble twin experiments. Noisy synthetic observations of albedo and snow depth, generated using the multilayer CryoGrid community model with a full energy balance, are assimilated using two ensemble-based data assimilation schemes: the particle batch smoother and the ensemble smoother. A comprehensive evaluation exercise demonstrates that the joint assimilation of albedo and snow depth improves the simulation skill by up to $86\%$ relative to the prior in specific glacier regions. The particle batch smoother excels in representing albedo dynamics, while the ensemble smoother is particularly effective for snow depth under low snowfall conditions. By combining the strengths of both observations, the joint assimilation achieves improved mass balance simulations across different glacier zones using either assimilation scheme. This work underscores the potential of ensemble-based data assimilation methods for refining glacier models by offering a robust framework to enhance predictive accuracy and reduce uncertainties in cryospheric simulations. Further advances in glacier data assimilation will be critical to better understanding the fate and role of Arctic glaciers in a changing climate.
\end{abstract}

\end{frontmatter}

\section{Introduction}

\noindent Glaciers are regarded as one of the key indicators of climate change. Over the past three decades, global glacier ice loss has contributed nearly 1 mm annually to sea level rise \citep{IPCC2022IPCC, Zemp2019Global2016}. At the same time, glaciers serve as a critical component of mountain water towers, helping to provide a more consistent and reliable water supply to downstream regions \citep{Immerzeel2019ImportanceTowers, Zhang2023OceanicTower}. Arctic glaciers are experiencing an accelerated mass loss \citep{Rounce2023GlobalMatters, VanPelt2019A1957-2018,stby2017Diagnosing1957-2014, Schmidt2023MeltwaterSvalbard} because warming is amplified in the Arctic at two to four times the global average through various positive feedback mechanisms \citep[e.g.][]{Rantanen2022The1979,Lind2018ArcticImport}. Freshwater runoff from melting Arctic glaciers can have considerable impacts on ocean circulation and ocean-atmosphere interaction globally \citep{Pontes2024WeakeningCentury}, as well as on regional marine biogeochemistry and productivity \citep{Ezat2024ArcticInterglacial, Hopwood2020ReviewArctic}. Thus, accurate knowledge of glacier mass balance is vital for understanding, detecting, and predicting the impacts of climate change. 

Glacier modeling is a primary method to determine the surface mass balance of glaciers, especially at sites with scarce long-term in situ observations, to reconstruct the past or project the future glacier evolution. Climate-driven glacier models include temperature-index models \citep[e.g.][]{Hock2003TemperatureAreas, Marzeion2012TheGlaciers} and energy balance models \citep[e.g.][]{Hock2005ASweden, Westermann2023TheCryosphere}. Temperature-index models approximate the melt rate based on air temperature \citep{Huss2015ARise}, while physically-based energy balance models explicitly calculate the energy fluxes on the glacier surface and therefore provide a more detailed representation of the processes controlling the surface mass balance. The accuracy of glacier models is limited by uncertainties related to meteorological forcing \citep{Marzeion2020PartitioningChange}, incomplete model physics \citep{Schmidt2023MeltwaterSvalbard}, and parameter uncertainty \citep{Rounce2020QuantifyingAsia}. Constraining each uncertainty source remains a significant challenge. 
 
Data assimilation methods can incorporate observations into modeling to improve accuracy and constrain simulation uncertainty \citep{Evensen2022DataFundamentals}. In situ and remotely sensed observations can be individually or jointly assimilated into glacier models, leading to a reduction of the aforementioned uncertainties \citep{Choi2023ImpactGreenland, Gillet-Chaulet2020AssimilationFilter}. The assimilation of ground-based glaciological measurements into glacier mass balance models is gradually becoming a recognized approach for updating glacier model parameters or initial states \citep{Landmann2021AssimilatingFilter, Sjursen2023BayesianUncertainties}. However, despite this recognition, there are still relatively few studies that have implemented data assimilation in glacier mass balance modeling. Moreover, in situ measurements are available for only a minority of glaciers worldwide, which presents a significant challenge to transfer information to the unmeasured majority of glaciers. In addition to the direct assimilation of mass balance measurements, other quantities that can directly influence mass balance changes, such as remotely sensed albedo or snow depth, can also be ingested within a data assimilation framework. 

Albedo, defined as the reflectivity of the Earth's surface to insolation, is a controlling variable in the surface energy balance of glaciers. It significantly impacts the shortwave radiation budget, thereby influencing the rate of melt and overall mass balance of glaciers \citep[e.g.][]{budyko1969,Ye2024UnveilingGlaciers}. In a pioneering study, a variational assimilation scheme was used to incorporate moderate resolution imaging spectroradiometer (MODIS) derived albedo into a snowpack model to reconstruct the spatial mass balance distribution for an Alpine glacier \citep{Dumont2012VariationalGlacier}. More recently, Sentinel-2 albedo estimates were assimilated into a glacio-hydrological model to improve the simulation of streamflow in two glacierized basins in the Canadian Rockies \citep{Bertoncini2024AssimilationHydrology}. 

Snowfall is another major driver of the mass balance, as it is the primary source of glacier mass gain \citep{Hock2003TemperatureAreas, Pramanik2019ComparisonPrecipitation}. Satellite-based snow depth retrievals, such as from the ICESat-2 laser altimeter, are a potentially globally available constraint on uncertainties in snowfall forcing which is being explored for seasonal snow data assimilation \citep{Mazzolini2024Spatio-temporalAltimeter}.  However, to the best of our knowledge, no experiment has explored the joint assimilation of remotely sensed albedo and snow depth into an energy balance model for glacier mass balance simulation. Moreover, the current state of the art in using Bayesian data assimilation to infer glacier surface mass balance has focused on static parameters in temperature index models using relatively costly Markov chain Monte Carlo methods \citep{Rounce2020QuantifyingAsia,Sjursen2023BayesianUncertainties}. This stands in contrast to other recent cryospheric work on glacier flow \citep{Brinkerhoff2022VariationalScale}, ice sheet \citep{Navari2021Reanalysis20002014}, seasonal snow \citep{Alonso-Gonzalez2022TheV1.0}, and permafrost \citep{Groenke2023InvestigatingTransfer}, which employ a greater diversity of modern Bayesian data assimilation (also known as inversion) schemes that allow for the use of more complex models.     

In this study, we performed twin experiments \citep{Arnold1986Observing-systemsFuture.,Masutani2010}, also known as synthetic experiments or Observing System Simulation Experiments, to explore the benefits of assimilating albedo and snow depth on glacier mass balance simulations. This allowed us to test the data assimilation workflow in a series of targeted experiments while avoiding challenges of real observations and model discrepancies \citep{Masutani2010}. In particular, as satellite-based measurements of albedo and snow depth and their associated error characteristics are not always available or consistent due to variable weather conditions and observational limitations, here synthetic observations are instead generated using synthetic truth (also known as nature) runs of the energy balance model CryoGrid \citep{Westermann2023TheCryosphere, Schmidt2023MeltwaterSvalbard}. These synthetic observations serve as idealized representations of satellite measurements, with their spatiotemporal resolution designed to mirror that of actual satellite data, forming the foundation for observing system simulation. To assimilate the synthetic observations, we employed and compared two Bayesian data assimilation schemes, namely the Particle Batch Smoother \citep[PBS;][]{Margulis2015} and the Ensemble Smoother \citep[ES;][]{vanLeeuwen1996}. The synthetic observations were derived from synthetic truth runs for four distinct scenarios, each representing different climatic conditions. These scenarios were selected to better capture the varying information content of the assimilated observations. The simulations were driven by reanalysis data from the Copernicus Arctic Regional Reanalysis (CARRA) dataset over 12 hydrological years from September 2010 to September 2022. Kongsvegen glacier, one of the best studied glaciers in High Arctic Svalbard, was selected as the study area due to availability of data and its extensive size, encompassing diverse glacier zones that offer a comprehensive basis for representing a broad range of Arctic glaciers. By conducting a large number of twin experiments, we compared the effectiveness of the particle-based PBS scheme to the ensemble Kalman-based ES in improving simulated glacier mass balance across different glacier zones. The sensitivity of the mass balance estimates to ensemble size was also tested in terms of both accuracy and precision using the PBS. Our methodology incorporated synthetic estimates of albedo and snow depth, along with gap-free surface mass balance values, enabling a robust evaluation of model performance in the absence of consistently reliable satellite-based observational data.

\section{Data and Methods}

\begin{figure*}
    \centering
    \hspace{5pt}
    \includegraphics[width = 150mm]{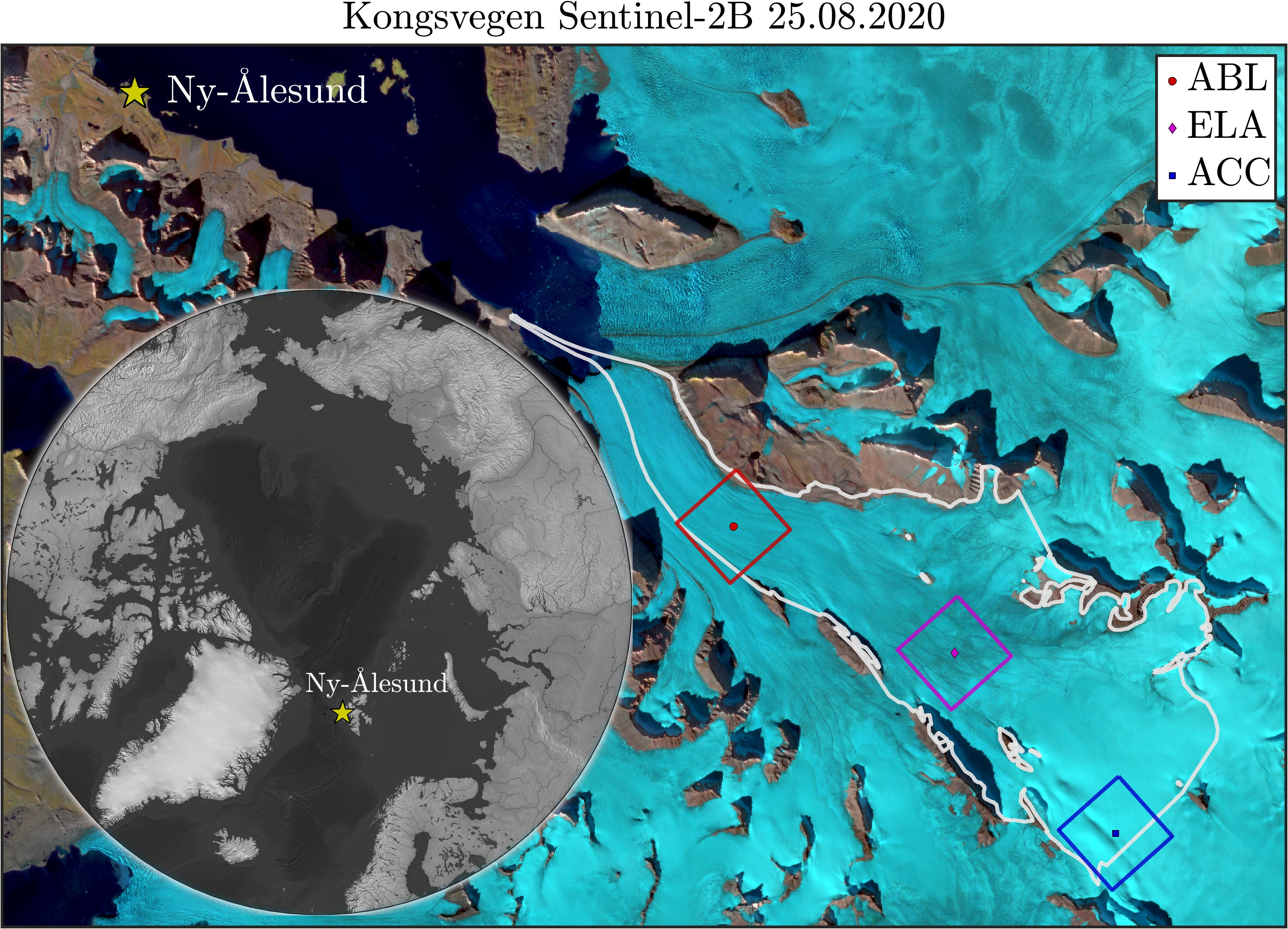}
    \caption{Atmospherically corrected shortwave infrared false color image over the area surrounding Kongsvegen glacier near Ny-\r{A}lesund in the Svalbard archipelago captured by the Sentinel-2B satellite at 13:07 UTC on the 25$^\mathrm{th}$ of August 2020. The image shows the locations of Ny-\r{A}lesund (yellow star) and the Kongsvegen glacier outline from RGI (white) as well as the locations of $2.5$ by $2.5$ km grid cells that were extracted from CARRA to represent the ablation zone (ABL, red), Equilibrium Line Altitude (ELA, purple), and the accumulation zone (ACC, blue) of Kongsvegen. The inset shows the location of Ny-\r{A}lesund (yellow star) in the Arctic (here roughly defined as latitudes above $60^\circ$N) on a polar stereographic map using open Gray Earth data from Natural Earth.}
    \label{fig:location}
\end{figure*}

\subsection{Study Site}
\noindent The Svalbard archipelago is one of the most climatically sensitive regions in the world \citep{Geyman2022Historical2100, Noel2020LowVariability}. For example, it is the region in Europe that has experienced the greatest warming in the past three decades \citep{Isaksen2016RecentCover, Nordli2014Long-term1898-2012}. Kongsvegen is located on the northwestern coast of Svalbard close to the research station of Ny-Ålesund (Fig.~\ref{fig:location}). The glacier has an area of around 100 km\textsuperscript{2} and a length of 26 km, with slopes ranging from 0.5 to 2.5\textdegree \citep{Karner2013ASvalbard}. The ice flows towards the northwest from its ice-divide at about 800 m a.s.l. down to sea level at the head of Kongsfjorden \citep{Karner2013ASvalbard, Hagen1999MassSvalbard}. The three grid cells used in this study are shown in Fig.~\ref{fig:location} and we used these grids to represent different glacier zones, namely the ablation area, equilibrium line altitude (ELA), and accumulation area. 

\subsection{Forcing Data}
This study uses the Copernicus Arctic Regional Reanalysis (CARRA) dataset \citep{CopernicusClimateChangeServiceC3SClimateDataStoreCDS2024ArcticCDS} as meteorological forcing data. The CARRA forcing fields considered are the 2 m air temperature, 2 m specific humidity, 10 m windspeed, incoming longwave and shortwave radiation, precipitation, and atmospheric pressure. CARRA is derived from the HARMONIE-AROME numerical weather prediction system \citep{Bengtsson2017TheSystem}. This regional reanalysis covers two domains in the European sector of the Arctic, CARRA-West and CARRA-East, employing ERA5 reanalysis as boundary conditions \citep{Yang2021SeaC3S}. The CARRA output has a horizontal resolution of $2.5$ km and a 3-hour temporal resolution covering the period from 1991 to present. Following \citet{Schmidt2023MeltwaterSvalbard}, this study employs meteorological data from the CARRA-East domain over 12 hydrological years from the 16$^\mathrm{th}$ of September 2010 to the 15$^\mathrm{th}$ of September 2022. 

\subsection{Mass balance model}
The CryoGrid community model is an open-source model developed for climate-driven multiphysics simulations of the terrestrial cryosphere \citep{Westermann2023TheCryosphere}, which uses a full surface energy-balance scheme that can be coupled to different multilayer subsurface modules of varying complexity. We used the glacier surface mass balance configuration of CryoGrid with a snow and firn module that was first employed by \citet{Schmidt2023MeltwaterSvalbard} and refer to this publication for more details. In this study, we added ensemble-based data assimilation methods, thereby creating a comprehensive probabilistic modeling package.

Through data assimilation, we aim to improve simulated glacier mass balance. In numerical weather forecasting and climate modeling, precipitation, particularly snowfall, is associated with significant uncertainties that can contribute to considerable errors in Arctic mass balance models \citep[e.g.][]{Forbes2011,Schmidt2017TheObservations,VanPelt2019A1957-2018,Lenaerts2020Present-dayModel}. To address these uncertainties, the CryoGrid model employs a relative bias correction for snowfall using a multiplicative snowfall factor $\beta_s$. 
Albedo is a controlling variable of the surface energy balance, and therefore accurate simulations are important for modeling the surface mass balance \citep[e.g][]{Schmidt2017TheObservations,Gunnarsson2023ModelingAlbedo}. In this study, snow albedo is calculated using the CROCUS snow spectral albedo scheme \citep{Vionnet2012TheV7.2}, where the albedo depends on the snow age, the optical grain diameter, and an albedo evolution rate $\tau_a$. Our ensemble assimilation approach involved constraining these uncertain parameters, $\beta_s$ and $\tau_a$, to enhance the accuracy of the simulated snowfall, the snow albedo evolution, and the associated surface mass balance. 

We initialized the model using a 5-year spin-up from 2006 to 2010, using a snow albedo evolution rate of $\tau_a=0.005$ day$^{-1}$ and no bias correction of the snowfall, i.e. $\beta_s=1$. This allows for appropriate representation of near-surface ice temperatures and the buildup of a small firn layer of $3$ m w.e. in the accumulation zone and improves the physical consistency of the subsequent experiments, particularly in the accumulation area. 
 
\subsection{Synthetic observations}
\noindent In this study, synthetic albedo and snow depth observations were assimilated to constrain simulations of glacier mass balance using the CryoGrid model. We generated albedo and snow depth time series for each of the three grid cells, selected to represent the ablation area, ELA, and accumulation area of Kongsvegen. By prescribing true parameters and running the model for these three grid cells, we obtained a synthetic truth from which synthetic observations have been generated. 

\begin{figure*}
    \centering
    \hspace{5pt}
    \includegraphics[width = 150mm]{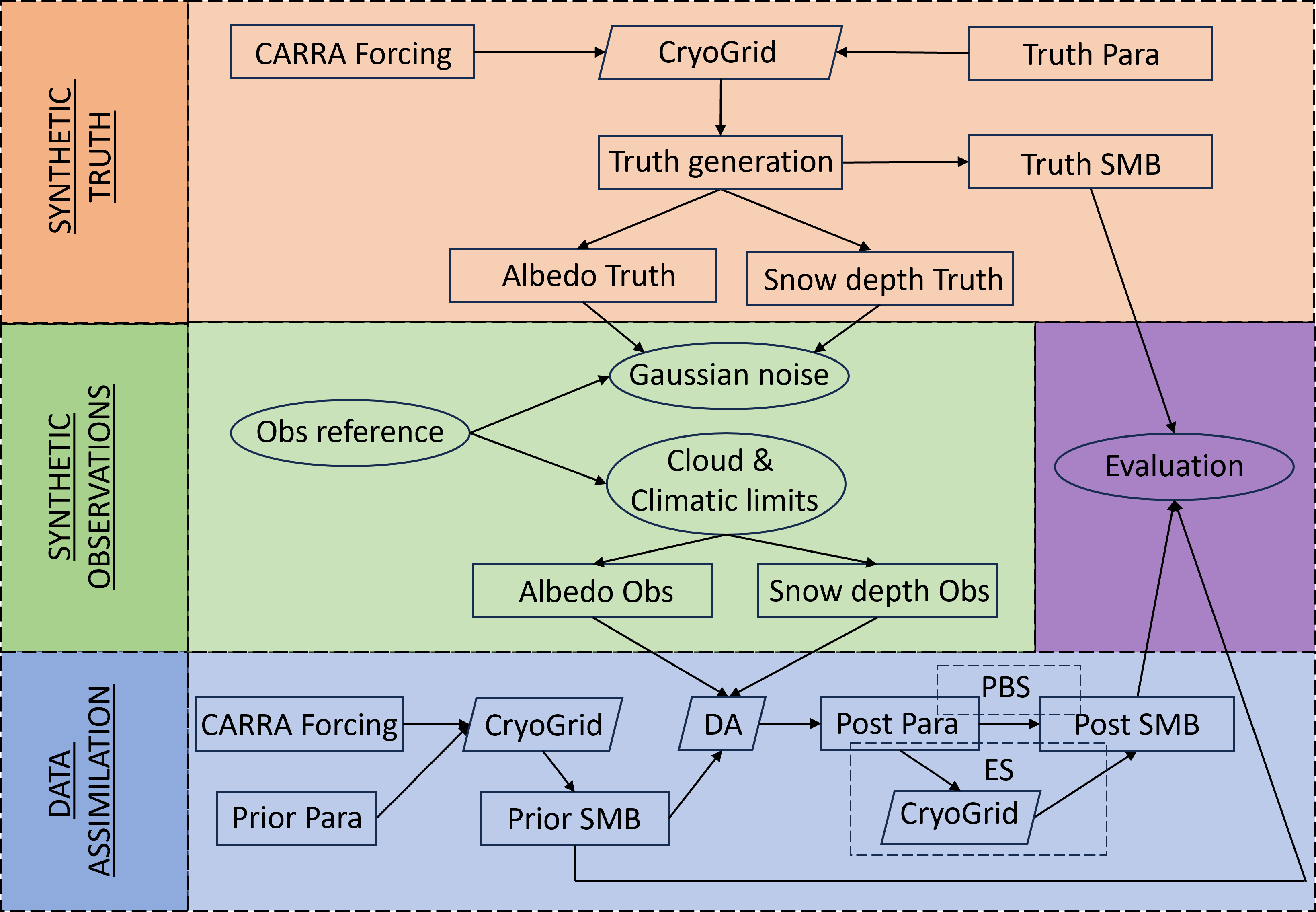}
    \caption{Workflow in the twin experiments involving the sequential generation of: synthetic truth runs (orange), noisy synthetic observations (green), and data assimilation experiments (blue) followed by the evaluation of each experiment (purple).}
    \label{workflow}
\end{figure*}

Fig. \ref{workflow} shows the workflow in our experimental design. We primarily divided the workflow into three steps. First, the generation of synthetic truth data was achieved by prescribing `true' parameters representing different conditions. To mimic realistic observational data that is inherently noisy, we added Gaussian noise to both the albedo and snow depth truth to represent observation error. For the albedo, this noise has a mean of 0 and an observation error standard deviation of $\sigma_\alpha=0.1$. This standard deviation is chosen based on the upper limit values of the reported root mean square error obtained by comparing MODIS albedo retrievals to in situ measurements \citep{Stroeve2005AccuracyMeasurements}. The noise added to snow depth has a mean of 0 m and a standard deviation of $\sigma_s=0.5$ m, which is based on the findings of \citet{Deschamps-Berger2023EvaluationData} for ICESat2 snow depth retrievals on low slopes. 

The synthetic albedo observations were sampled based on the effective temporal resolution of MODIS onboard the Terra and Aqua satellites. We simulated the impact of polar night on the availability of optical albedo retrievals in our research area by removing the values of synthetic albedo during this time (November to February). In addition, the occluding impact of cloud cover is considered in this study. According to findings by \cite{Marshall1993LimitationsRegions}, statistically only 22\% to 24\% of days between April and September in Svalbard are classified as clear-sky conditions, making MODIS albedo products usable only for those days. Also, \cite{stby2014SevereSvalbard} found that only 26\% of MODIS products are acquired under clear sky conditions on Austfonna, Svalbard. Thus, we use 30 daily albedo observations that were randomly distributed in time excluding polar night, representing approximately 20\% of the total days of each year between mid of April to mid of October. In our analysis of snow depth data, we considered the temporal resolution provided by the ICESat-2 satellite in the Arctic region. Typically, ICESat-2 operates on a 91-day revisit cycle at the equator. However, due to its high-inclination orbit, the ground tracks of the satellite converge towards the poles, significantly enhancing the frequency of overpasses in polar regions. Consequently, in the Arctic, the temporal resolution increases, with revisit intervals reduced to approximately 1 to 2 weeks \citep{Markus2017TheImplementation}. This enhancement in revisit frequency was utilized to simulate the temporal resolution in our synthetic snow depth data, providing a more accurate representation of snow accumulation and change over time in this region.

\subsection{Data assimilation}
\label{dae}
\noindent In this section we describe the data assimilation methods used and their implementation in CryoGrid to infer glacier mass balance. 

For a more comprehensive treatment of Bayesian data assimilation methods we refer to the extensive work of \citet{Evensen2022DataFundamentals} and \citet{Sanz-Alonso_Stuart_Taeb_2023} and the overview in \citet{Alonso-Gonzalez2022TheV1.0} for details pertinent to cryospheric applications. Data assimilation is loosely defined as the fusion of data and models that can be mathematically formalized using the probabilistic framework of Bayesian inference. 

There are multiple sources of model uncertainty related to the choice of model parameters, forcing, initial conditions, and model structure. Herein, we are primarily concerned with the two first sources of uncertainty which we lump into an uncertain input parameter vector $\boldsymbol{\theta}$ with $N_p=2$ elements, namely $\beta_s$ and $\tau_a$. By using the synthetic truth generated by the same underlying model, CryoGrid in this case, we avoid model structural uncertainty by confining ourselves to so-called identical twin experiments \citep{Arnold1986Observing-systemsFuture.}. 

Within these identical twin experiments, we are thus by construction justified in restricting ourselves to solving the strong constraint data assimilation problem that assumes a perfect data generating model (CryoGrid) that can map perfectly onto reality if the true input vector $\boldsymbol{\theta}^\star$ were known \citep{Evensen2022DataFundamentals}. Note that strong constraint problems are often also solved in practice in real experiments where the underlying perfect model assumption is always violated to some extent \citep[e.g.][]{Alonso-Gonzalez2022TheV1.0}.  

In this strong constraint setting, we can model a vector of $N_o$ noisy observations $\mathbf{y}$ according to the following data generating process
\begin{equation}
\mathbf{y}=\mathcal{G}\left(\boldsymbol{\theta}^\star\right)+\boldsymbol{\epsilon} \, . 
\label{eq:forward}
\end{equation}
where $\mathcal{G}(\cdot)$ denotes the data generating model, $\boldsymbol{\theta}^\star$ is the aforementioned true parameter vector, and $\boldsymbol{\epsilon}$ is a noise term representing observation error. Given some observations, the task at hand is to invert $\mathcal{G}(\cdot)$ to recover $\boldsymbol{\theta}^\star$. This task is challenging since $\mathcal{G}(\cdot)$ is often a nonlinear and relatively computationally costly model instantiated in a long piece of typically non-differentiable code, namely CryoGrid in our case \citep{Westermann2023TheCryosphere}. To complicate matters further, it is also a fundamentally ill-posed inverse problem \citep{Sanz-Alonso_Stuart_Taeb_2023}.

Adopting a probabilistic perspective naturally leads to casting this ill-posed inverse problem in terms of Bayesian inference \citep{Sanz-Alonso_Stuart_Taeb_2023}, and the computational challenge motivates the adoption of efficient ensemble-based data assimilation algorithms to make inference tractable \citep{Evensen2022DataFundamentals}. Formally the entire exercise of data assimilation can now be boiled down to using Bayes' rule as follows
\begin{equation}
p(\boldsymbol{\theta}|\mathbf{y})=\frac{p(\mathbf{y}|\boldsymbol{\theta})p(\boldsymbol{\theta})}{p(\mathbf{y})} \label{eq:Bayes} \, ,
\end{equation}
to infer the posterior probability distribution $p(\boldsymbol{\theta}|\mathbf{y})$ over parameters $\boldsymbol{\theta}$ given data $\mathbf{y}$. The likelihood quantifies how well the model predictions with parameters $\boldsymbol{\theta}$ fit the noisy observations $\mathbf{y}$, the prior regularizes the problem using background information about $\boldsymbol{\theta}$, and the evidence $p(\mathbf{y})$ is a normalizing constant \citep{MacKay2003}. 

Once prior and likelihood are defined, Bayesian inference is theoretically straightforward and is just a matter of applying \eqref{eq:Bayes} to a grid of parameter vectors $\boldsymbol{\theta}$. Practical geophysical applications of Bayesian inference for data assimilation tend to require more efficient methods than computationally expensive grid approximations. The current state-of-the-art data assimilation approaches can generally be split into ensemble-based (Monte Carlo) and variational methods \citep{Evensen2022DataFundamentals}. The latter requires a differentiable model which is often, as is the case with this CryoGrid version \citep{Westermann2023TheCryosphere}, not available. As such, we use ensemble-based data assimilation methods that are widely used in cryospheric applications \citep[e.g.][]{Navari2021Reanalysis20002014, Alonso-Gonzalez2022TheV1.0, Groenke2023InvestigatingTransfer}, but have not widely applied much to glacier mass balance. In particular, we adopt both the PBS and the ES to compare their performance for a large ensemble of twin experiments. In addition to being used in the literature \citep{Alonso-Gonzalez2022TheV1.0}, these methods are relatively straightforward to implement and can serve as kernels for more sophisticated schemes. 

\subsubsection{Prior and likelihood}

In this study, we focus on two uncertain parameters within the glacier configuration of CryoGrid, namely the albedo evolution rate $\tau_a$ and the snowfall factor $\beta_s$. The former factor $\tau_a$ is an inverse timescale that controls the rate at which the visible albedo in the Crocus albedo parametrization decays \citep{Vionnet2012TheV7.2}, with larger (smaller) values indicating a faster (slower) decay rate. The latter multiplier $\beta_s$ explicitly accounts for biases in the snowfall (solid precipitation) forcing from the CARRA reanalysis while also implicitly accounting for unresolved processes in this instantiation of CryoGrid in the form of wind-driven snow redistribution. Both parameters are treated as fixed (time-invariant) within a given mass balance year. As such, in this study the parameter vector $\boldsymbol{\theta}$ has $N_p=2$ elements so we consider a 2D parameter space. On the one hand, this is quite a low-dimensional parameter space. On the other hand, CryoGrid which we use as the data generating model is relatively expensive to evaluate. Moreover, these $N_p=2$ parameters were selected based on several modeling studies of glacier mass balance where these were deemed among the most uncertain yet sensitive parameters \citep[e.g.][]{Schmidt2017TheObservations,VanPelt2019A1957-2018,Lenaerts2019,Raoult2023,Schmidt2023MeltwaterSvalbard}.

To encode uncertainty in these parameters we need to specify a prior distribution $p(\boldsymbol{\theta})$ that reflects our prior knowledge concerning possible values for these parameters. Herein, building on several related studies \citep{Aalstad2018Ensemble-basedSites,Mazzolini2024Spatio-temporalAltimeter,Guidicelli2024,Keetz2024}, we use the generalized logit-normal prior distribution that is a double bounded transformed version of a normal distribution allowing for upper and lower bounds $(a,b)$, a central location parameter $\mu_0^*$, and a scale parameter $\sigma_0$ reflecting the spread in possible values. Following \citet{Keetz2024}, this prior is defined as follows for a scalar parameter $\theta$ % 
\begin{equation}
    p(\theta|\mu_0,\sigma_0,a,b)=\frac{|J|}{\sigma_0\sqrt{2\pi}}\exp\left(-\frac{\left(\phi-\mu_0\right)^2}{2\sigma_0^2}\right) \, , \label{eq:prior}
\end{equation}
where $|J|=(b-a)/(\theta-a)(b-\theta)$ is a Jacobian term and $\phi$ is the generalized logit transform of $\theta$
\begin{equation}
    \phi=\psi(\theta,a,b)=\ln\left(\frac{\theta-a}{b-a}\right)-\ln\left(\frac{b-\theta}{b-a}\right) \, , \label{eq:anamorphosis}
\end{equation}
with corresponding inverse transform 
\begin{equation}
\theta=\psi^{-1}(\phi,a,b)=a+\frac{b-a}{1+\exp(-\phi)} \, .\label{eq:invana}
\end{equation}
The generalized logit normal distribution in \eqref{eq:prior} has an associated normal distribution, namely the distribution of the logit transformed parameter $\phi$ with mean $\mu_0$ and standard deviation $\sigma_0$. We use the $\sigma_0$ parameter to define the scale (spread) of the logit-normal prior for a bounded parameter $\theta$. For the location parameter of the logit normal we use the median of the logit normal distribution $\mu_0^*$ that can be transformed to the mean of the associated normal distribution through $\mu_0=\psi(\mu_0^*,a,b)$ and vice versa. 

The generalized logit prior in \eqref{eq:prior} is specified independently for each of the parameters $\tau_a$ and $\beta_s$ using the hyperparameters in Table~\ref{tab:prior}. Assuming independence, the joint prior $p(\boldsymbol{\theta})$ is simply given by the product of the marginal priors $p(\boldsymbol{\theta})=p(\tau_a)p(\beta_s)$. 
To sample from the generalized logit normal distribution \eqref{eq:prior}, we apply the generalized logit transform \eqref{eq:anamorphosis} to the prior median $\mu_0^*$ to obtain the mean of the associated normal $\mu_0$ then add $N_e$ samples of randomly generated Gaussian noise with standard deviation $\sigma_0$ to $\mu_0$ and apply the inverse transform \eqref{eq:invana} to obtain prior samples $\theta_i\sim p(\theta)$ for the parameter $\theta\in\boldsymbol{\theta}$ (i.e., either $\tau_a$ or $\beta_s$) in question. After having done this for both parameters, we are left with an ensemble of $N_e$ particles from the joint prior $\boldsymbol{\theta}_i\sim p(\boldsymbol{\theta})$. 

\begin{table*}
\caption{The hyperparameters for the independent logit-normal priors used for each of the $N_p=2$ uncertain parameters in the parameter vector $\boldsymbol{\theta}$ considered in this study. The hyperparameters are the lower bound $a$, the upper bound $b$, the location parameter which is the median $\mu_0^*$, and the scale (spread) which is the dimensionless standard deviation $\sigma_0$ of the associated normal distribution.}
\centering
\begin{tabular}{ccccccc}
\hline
Parameter name & Symbol & Units & Lower bound $a$ & Upper bound $b$ & Location $\mu_0^*$ & Scale $\sigma_0$ \\
\hline
Albedo evolution rate & $\tau_a$ & $\mathrm{day}^{-1}$  & $0.0001$ & $0.05$  & $0.005$  & $1$ \\
Snowfall factor & $\beta_s$ & - & $0.5$ & $2$ & $1$ & $1$ \\
\hline
\end{tabular}
\label{tab:prior}
\end{table*}

As is commonly done in data assimilation \citep{Carrassi2018}, we use a simple additive zero-mean Gaussian observation error model of the form $\boldsymbol{\epsilon}\sim\mathrm{N}(\mathbf{0},\mathbf{R})$ where $\mathbf{R}$ is an $N_o\times N_o$ observation error covariance matrix. This can be justified as a useful default first-order error model using both the central limit theorem and maximum entropy arguments \citep{Jaynes2003}. Using this error model, allows us to formulate the likelihood $p(\mathbf{y}|\boldsymbol{\theta})$. By definition, this is the probability density of the (fixed) observations $\mathbf{y}$ given that the parameter set $\boldsymbol{\theta}$ is true. By inspection of \eqref{eq:forward} conditional on $\boldsymbol{\theta}=\boldsymbol{\theta}^\star$ the observation error becomes $\boldsymbol{\epsilon}=\mathbf{y}-\mathcal{G}(\boldsymbol{\theta})$ and by inserting this into the Gaussian observation error model we obtain a Gaussian likelihood $p(\mathbf{y}|\boldsymbol{\theta})=\mathrm{N}(\mathbf{y}|\mathcal{G}(\boldsymbol{\theta}),\mathbf{R})$ of the form
\begin{equation}
    p(\mathbf{y}|\boldsymbol{\theta})=c_y \text{exp}\left(-\frac{1}{2}\left[\mathbf{y}-\widehat{\mathbf{y}}\right]^\mathrm{T}\mathbf{R}^{-1}\left[\mathbf{y}-\widehat{\mathbf{y}}\right]\right) \, , \label{eq:likelihood}
\end{equation}
where $c_y=\det(2\pi\mathbf{R})^{-1/2}$ is a constant and $\widehat{\mathbf{y}}=\mathcal{G}(\boldsymbol{\theta})$ denotes the predicted observations from the data generating model given a particular parameter set $\boldsymbol{\theta}$. Following the likelihood principle in Bayesian inference \citep{Jaynes2003}, the likelihood should be viewed as a function of the uncertain parameters $\boldsymbol{\theta}$ (here through $\widehat{\mathbf{y}}=\mathcal{G}(\boldsymbol{\theta})$) rather than a distribution over the fixed (albeit noisy) observations $\mathbf{y}$ that we are assimilating. Although the likelihood in \eqref{eq:likelihood} is Gaussian, our data generating model $\widehat{\mathbf{y}}=\mathcal{G}(\boldsymbol{\theta})$ makes it nonlinear. 

To further simplify the likelihood \eqref{eq:likelihood} we also make a standard assumption that the observation errors are conditionally independent \citep{Carrassi2018,Sarkka2023}.  As such, our $N_o\times N_o$ observation error covariance matrix $\mathbf{R}$ becomes diagonal with entries corresponding to the observation error variance $\sigma_{y_m}^2$ associated with each of the $m=1,\dots,N_o$ observations $y_m$ in the observation vector $\mathbf{y}$. When we only assimilate one type of observation, these entries are constant and equal to the observation error variance of either snow depth ($\sigma_d^2$) or albedo ($\sigma_\alpha^2$). For joint assimilation, where both types of observation are assimilated, both error variances appear along the diagonal of $\mathbf{R}$ in accordance with the entries in $\mathbf{y}$. 

\subsubsection{Particle batch smoother}
The Particle Batch Smoother was introduced in the snow literature by \citet{Margulis2015} as a batch smoother version of the widely used particle filter \citep[see][]{Chopin2020,Sarkka2023}. Algorithmically, the PBS boils down to performing basic sequential importance sampling \citep{vanLeeuwen2009} which effectively represents the posterior through a particle approximation
\begin{equation}
p(\boldsymbol{\theta}|\mathbf{y})\simeq \sum_{i=1}^{N_e} w_i \delta\left(\boldsymbol{\theta}-\boldsymbol{\theta}_i\right) \, , \label{eq:particleapp}
\end{equation}
where $w_i$ are the weights associated with each of the $i=1,\dots,N_e$ particles (samples) $\boldsymbol{\theta}_i$ in parameter space. These particles weights are self-normalized such that $\sum_{i=1}^{N_e}w_i=1$. The $\delta(\cdot)$ in \eqref{eq:particleapp} denotes the Dirac delta which is a generalized function with properties $\int \delta(\boldsymbol{\theta}-\boldsymbol{\theta}_i) \, \mathrm{d}\boldsymbol{\theta}=1$ and $\int g(\boldsymbol{\theta})\delta(\boldsymbol{\theta}-\boldsymbol{\theta}_i) \, \mathrm{d}\boldsymbol{\theta}=g(\boldsymbol{\theta}_i)$ for some function of the parameters $g(\boldsymbol{\theta})$. Thereby, the particle approximation represents the continuous posterior probability density function as a sum of discrete particles with probability mass given by their weights $w_i$. Posterior expectations become straightforwards to compute, for example setting $g(\boldsymbol{\theta})=\boldsymbol{\theta}$ we recover the particle approximation to the posterior mean of the parameters as the weighted sum over particles $\boldsymbol{\theta}_i$. The corresponding posterior expectations in state space are obtained analogously. With minimal loss of accuracy, simpler unweighted posterior statistics are computed by first resampling particles based on the weights \citep{Alonso-Gonzalez2022TheV1.0}. The weights $w_i$ in the PBS are obtained through basic importance sampling approach using the prior as a proposal distribution to sample particles $\boldsymbol{\theta}_i\sim p(\boldsymbol{\theta})$ so that the weights effectively become the likelihood ratio
\begin{equation}
    w_i=\frac{p(\mathbf{y}|\boldsymbol{\theta}_i)}{\sum_{k=1}^{N_e}p(\mathbf{y}|\boldsymbol{\theta}_k)} \, ,
\end{equation}
which when we insert for our Gaussian likelihood simplifies to \citep{Aalstad2018Ensemble-basedSites}
\begin{equation}
    w_i=\frac{\exp\left(-\frac{1}{2}\left[\mathbf{y}-\widehat{\mathbf{y}}_i\right]^\mathrm{T}\mathbf{R}^{-1}\left[\mathbf{y}-\widehat{\mathbf{y}}_i\right]\right)}{\sum_{k=1}^{N_e}\exp\left(-\frac{1}{2}\left[\mathbf{y}-\widehat{\mathbf{y}}_k\right]^\mathrm{T}\mathbf{R}^{-1}\left[\mathbf{y}-\widehat{\mathbf{y}}_k\right]\right)} \, , \label{eq:weights}
\end{equation}
where $\mathbf{y}_i=\mathcal{G}(\boldsymbol{\theta}_i)$ denotes the vector of $N_o$ predicted observables from CryoGrid for particle $i$ with associated parameter vector $\boldsymbol{\theta}_i$, $\left[\cdot\right]^\mathrm{T}$ denotes the transpose, and $\mathbf{R}^{-1}$ is the inverse of the $m\times m$ observation error covariance matrix. In practice, we first compute the logarithm of the PBS weights in \eqref{eq:weights} to ensure numerical stability as described in \citet{Alonso-Gonzalez2022TheV1.0}. Both the PBS and ES are batch smoothers in the sense that they assimilate a single batch of observations in a long data assimilation window, unlike a filter which updates sequentially as observations become available. The length of the window is typically defined by a typical timescale of the system being modeled, which we here take to be one mass balance year. This smoothing property is crucial since it allows the future to update the past: observations in the accumulation season can inform model states in the \emph{preceding} accumulation season \citep{Margulis2015,Aalstad2018Ensemble-basedSites}. A computational advantage of the PBS is that it only requires running a single ensemble model integration of $N_e$ particles sampled from the parameter prior. A particle approximation of the posterior for model parameters and state variables can then be obtained solely using the weights in \eqref{eq:weights} followed by a resampling step. As such, the computational cost of the PBS is incurred almost entirely by the need to run $N_e$ forward simulations of the data generating model $\mathcal{G}$. It is this feature that helped motivate our design of a large ensemble of twin experiments, in that it is straightforward to test a large number of observation types and parameter scenarios based on a single large ensemble run by using a (fixed) prior distribution $p(\boldsymbol{\theta})$ as the proposal.  

\subsubsection{Ensemble smoother}
We also test the ensemble smoother (ES) scheme that was originally proposed by \citet{vanLeeuwen1996} as a batch smoother version of the widely used ensemble Kalman filter \citep[EnKF][]{Evensen2022DataFundamentals}. Here we use the classic stochastic version of the ES with perturbed observations to avoid underestimating ensemble covariances \citep{vanLeeuwen2020}. The general framework of ensemble Kalman methods, which the ES falls under, extends the domain of applicability of classical Kalman filtering methods \citep{Sarkka2023}, that require Gaussian linear data generating models, to Gaussian nonlinear models \citep{Evensen2022DataFundamentals}. The Gaussian assumption in the prior and likelihood can also be relaxed through transformations using Gaussian anamorphosis functions \citep{Bertino2003}.  Herein we use an analytical approach to Gaussian anamorphosis using the generalized logit transform in \eqref{eq:anamorphosis}. 
Among the ensemble Kalman methods, the ES is most widely used for parameter estimation such as the strong constraint problem that we are tackling here. 

The ES is initialized by sampling an ensemble of $N_e$ parameter vectors $\boldsymbol{\theta}_i^{(0)}$ from the prior $\boldsymbol{\theta}_i^{(0)}\sim p(\boldsymbol{\theta})$. Using this prior parameter ensemble, following \citet{Aalstad2018Ensemble-basedSites} the stochastic ES with analytical anamorphosis proceeds in the following steps while looping over ensemble members $i=1,\dots,N_e$:
\begin{enumerate}
\item Generate an ensemble of prior predicted observables by running the parameters through the data generating model $\widehat{\mathbf{y}}_i^{(0)}=\mathcal{G}\left(\boldsymbol{\theta}_i^{(0)}\right)$ which implicitly also involves generating an ensemble of prior model state vectors $\mathbf{x}_i^{(0)}$ for the whole data assimilation window (i.e., mass balance year in our case). 
\item Transform the prior parameter ensemble to Gaussian space using Gaussian anamorphosis $\boldsymbol{\phi}_i^{(0)}=\boldsymbol{\Psi}(\boldsymbol{\theta}_i^{(0)})$ in the form of the generalized logit transform \eqref{eq:anamorphosis}.
\item Perform the ensemble Kalman analysis step to update the parameters
\begin{equation}
\boldsymbol{\phi}_i^{(1)}=\boldsymbol{\phi}_i^{(0)}+\mathbf{K}^{(0)}\left(\mathbf{y}+\boldsymbol{\epsilon}_i-\widehat{\mathbf{y}}_i^{(0)}\right) \label{eq:Kalman}
\end{equation}
where the ensemble Kalman gain $\mathbf{K}^{(0)}$ is obtained using ensemble covariance matrices together with $\mathbf{R}$ as outlined in \citet{Aalstad2018Ensemble-basedSites} while realizations of Gaussian observation noise $\boldsymbol{\epsilon}_i\sim\mathrm{N}(0,\mathbf{R})$ are used to perturb the observations $\mathbf{y}$ in this stochastic scheme \citep{vanLeeuwen2020}. 
\item Apply the inverse transformations using \eqref{eq:invana} to recover the posterior parameter ensemble in the original model parameter space $\boldsymbol{\theta}_i^{(1)}=\boldsymbol{\Psi}^{-1}(\boldsymbol{\phi}_i^{(1)})$.
\item Rerun the data generating model to obtain an ensemble of posterior predicted observables $\widehat{\mathbf{y}}_i^{(1)}=\mathcal{G}\left(\boldsymbol{\theta}_i^{(1)}\right)$ which also implicitly yields an ensemble of posterior model state vectors $\mathbf{x}_i^{(1)}$.
\end{enumerate}
Note that the parameters $\boldsymbol{\theta}$ are updated directly while the model state $\mathbf{x}$ are updated indirectly. 

As such, to recover the posterior state $\mathbf{x}$ with the ES it is necessary to run the data generating model twice for each ensemble member, first with the prior parameters $\boldsymbol{\theta}_i^{(0)}$ in step 1 and subsequently with the posterior parameters $\boldsymbol{\theta}_i^{(1)}$ in step 5. Thereby, for posterior state estimation the ES is twice as costly as the PBS in that it requires running the data generating model $2N_e$ times.

\subsection{Twin experiments}

\begin{figure*}
    \centering
    \hspace{5pt}
    \includegraphics[width = 150mm]{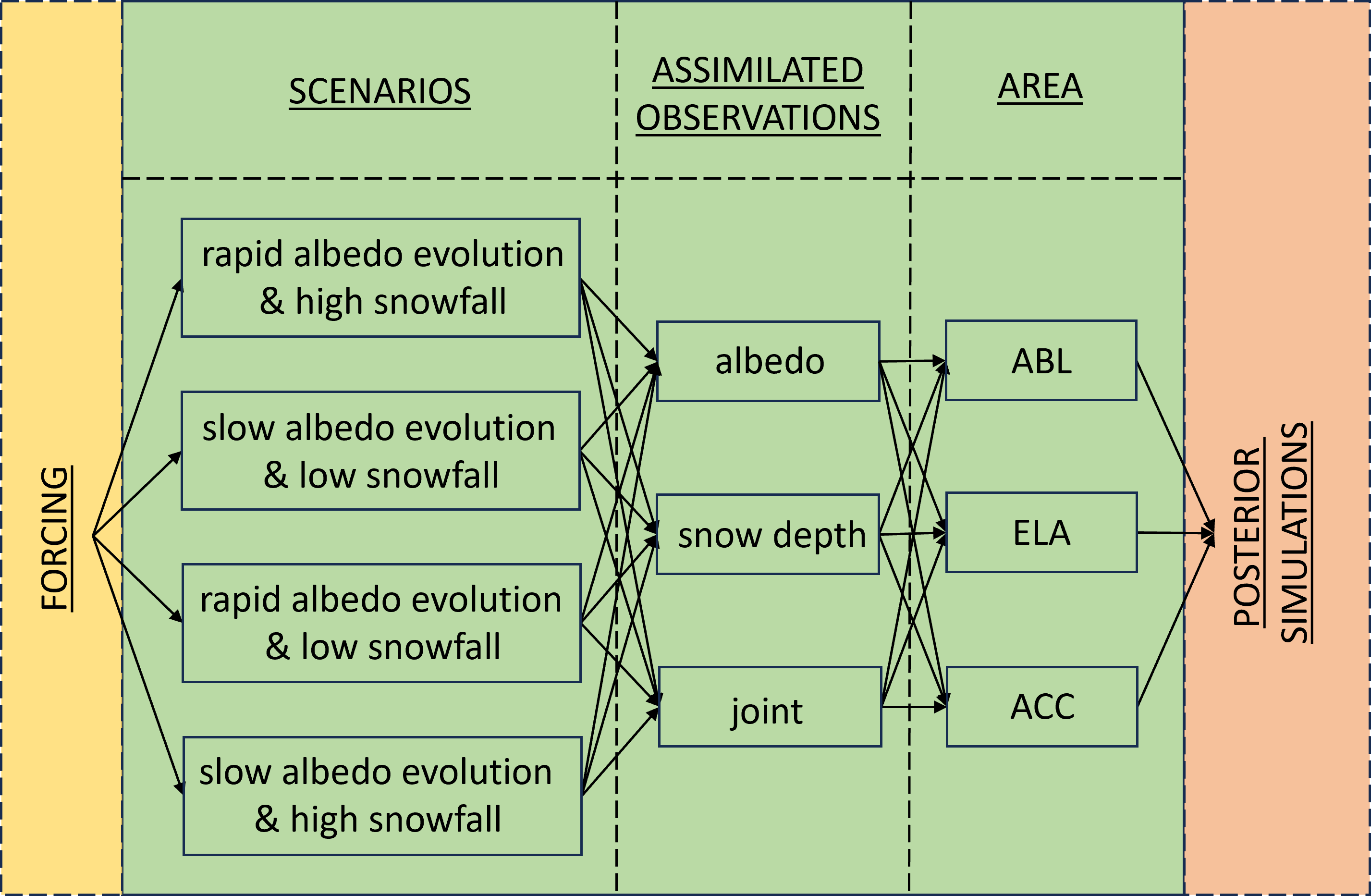}
    \caption{Structure of the large ensemble twin experiments based on permutations of four parameter scenarios, three types of assimilated observation vectors, and three experimental areas generating a total of $36$ twin experiments. The scenarios combine either a rapid or slow albedo evolution rate with either a high  or low snowfall factor. The assimilated observation vectors are either albedo only, snow depth only, or joint assimilation of albedo and snow depth. The experimental areas are either the ablation (ABL), equilibrium line altitude (ELA), and accumulation (ACC) areas depicted in Fig.~\ref{fig:location}.}
    \label{Scenarios}
\end{figure*}

The conceptual diagram in Fig.~\ref{Scenarios} shows the structure of the twin experiments where we generated synthetic truth scenarios along with synthetic noisy observations. We constructed four different scenarios (Fig.~\ref{workflow}) by using different true parameter vectors $\boldsymbol{\theta}^\star$ with different values for the true snow albedo evolution rate $\tau_a^\star$ and true snowfall factor $\beta_s^\star$. These true parameter vector scenarios include combinations of high and low values for each of the two parameters.  The true parameter vector scenarios are then used in CryoGrid to generate synthetic true state $\mathbf{x}^\star$ scenarios including the true observables $\widehat{\mathbf{y}}^\star=\mathcal{G}(\boldsymbol{\theta}^\star)$. These diverse true parameter scenarios are used to effectively mimic the variability of meteorological conditions and location-specific characteristics under different climatic scenarios. The synthetic albedo and snow depth obtained under these four different climatic scenarios were generated and perturbed with Gaussian noise that was scaled with the appropriate variances ($\sigma_{\alpha}^2$ and $\sigma_d^2$) to mimic observation error. These noisy synthetic observations are then assimilated to constrain the prior CryoGrid simulations. Note that in this assimilation exercise, the model has no access to the hidden synthetic truth ($\boldsymbol{\theta}^\star$, $\mathbf{x}^\star$, $\widehat{y}^\star)$ other than through the corrupted information present in the noisy synthetic observations. This is the standard setup for widely used identical twin experiments where the same model is used to generate the observations and in the subsequent assimilation experiments \citep{Arnold1986Observing-systemsFuture., Masutani2010}. The generated synthetic observations, albedo and snow depth, can be assimilated either individually or jointly, amounting to a total of  three assimilated observation scenarios. All experiments are applied in three different glacier zones, namely the ablation, ELA, and accumulation areas.

The prior ensemble of CryoGrid simulations consists of $N_e=1000$ ensemble members that were generated by perturbing albedo evolution rate and snowfall factor. We implemented the prior simulation with the same initial conditions for all data assimilation experiments. When initiating the model, we performed a 5 years spin-up to eliminate initialization shocks. Two different ensemble-based data assimilation methods, the PBS and ES, were compared in the twin experiments.

\subsection{Evaluation of the experiments}

To evaluate the performance of all experiments, we use the Continuous Ranked Probability Score (CRPS) to compare the posterior mass balance distribution to the synthetic truth mass balance. As outlined in \citet{Hersbach2000}, the CRPS is a statistical metric that compares probabilistic ensemble predictions to deterministic ground-truth values. Compared with Root Mean Square Error (RMSE) which is mainly used for deterministic forecasts, the CRPS is designed for probabilistic forecasts, which can evaluate the entire predictive distribution and provide a comprehensive assessment of the quality of predictions that include uncertainty quantification. The CRPS evaluates both the accuracy and the precision of the ensemble. The latter precision is a gauge of how well calibrated the ensemble is by punishing ensembles that are overconfident (too narrow) and underconfident (too wide). The CRPS is a negatively oriented score where a score of zero means that the probabilistic prediction is perfect, which only occurs for deterministic forecasts centered on the truth, while a larger CRPS entails a worse score. The CRPS is given by \citep{Gneiting2005}

\begin{equation}
\mathrm{CRPS}(P, x) = \int_{-\infty}^{\infty} \left( P(x) - H(x-x^\star) \right)^2 \, \mathrm{d}x
\end{equation}

\noindent where $P(x)$ is the cumulative distribution function of the ensemble prediction for variable $x$, $x^\star$ is the reference value which can be a synthetic truth or an observation, and $H(x-x^\star)$ is the Heaviside function, which is $1$ if $x \geq x^\star$ and $0$ otherwise. The CRPS inherits the same units as the variable $x$ whose ensemble prediction is being evaluated. 

\section{Results}

\subsection{Influence of observations on mass balance modeling by PBS}

\begin{figure*}
    \centering
    \hspace*{-0.5cm}
    \includegraphics[width = 178mm]{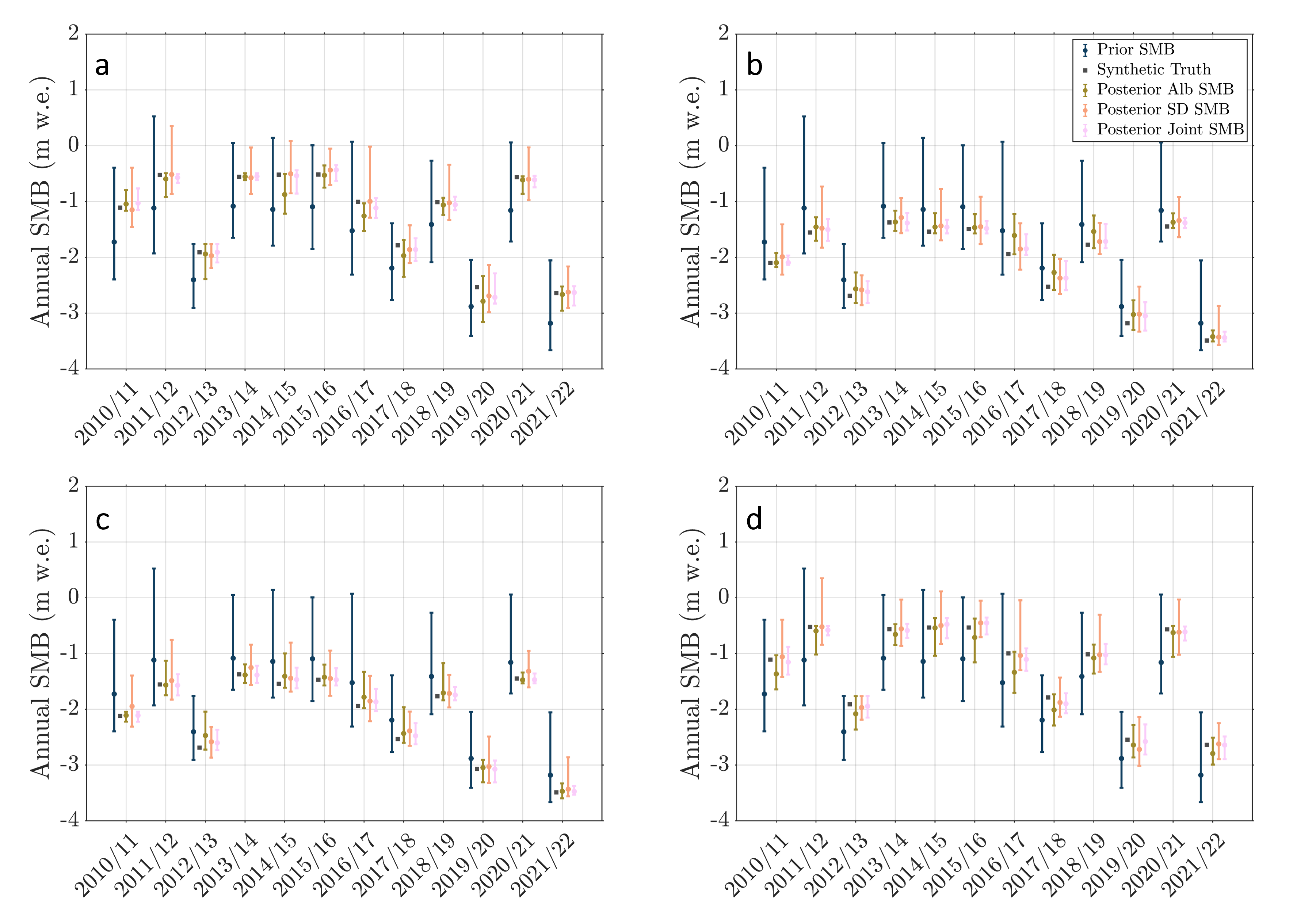}
    \caption{Comparison of prior, posterior, and true surface mass balance in the ablation area when using the PBS to assimilate albedo only, snow depth only, and both observations jointly. The figure presents four scenarios based on the snow albedo evolution rates and snowfall factors: a) Rapid snow albedo evolution with high snowfall. b) Slow snow albedo evolution with low snowfall. c) Rapid snow albedo evolution with low snowfall. d) Slow snow albedo evolution with high snowfall. Error bars represent the 95$^\text{th}$ central percentile range of the ensemble with the points indicating the median value for mass balance estimates.}
    \label{fig:PBS_abl}
\end{figure*}

\begin{figure*}
    \centering
    \hspace*{-0.5cm}
    \includegraphics[width = 178mm]{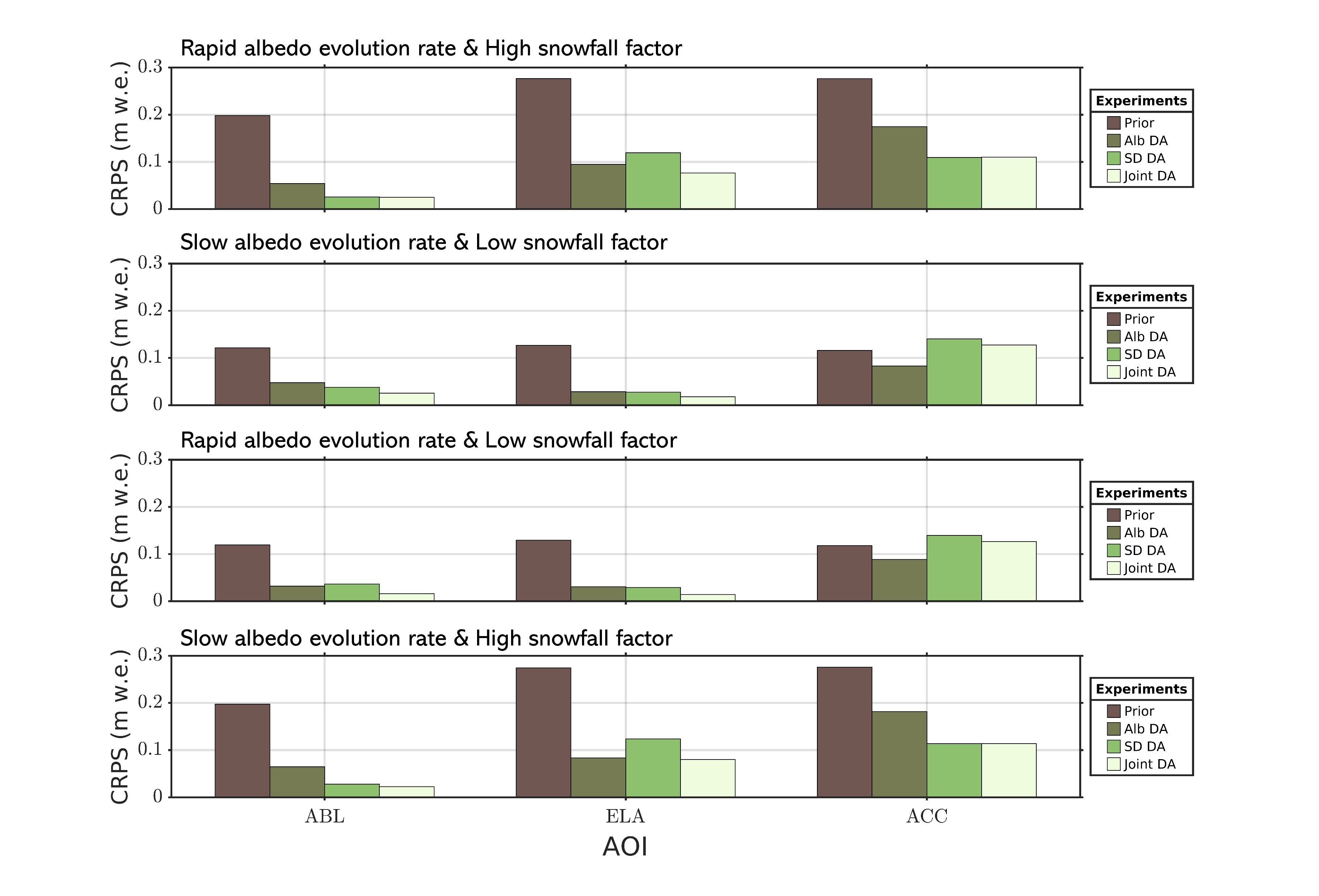}
    \caption{Continuous ranked probability score (CRPS) for the prior and posterior mass balance after assimilating albedo, snow depth, and both observations jointly using the PBS under all scenarios, compared to synthetic true mass balance.}
    \label{fig:CRPS_all}
\end{figure*}

Here, we present the results of surface mass balance (SMB) simulations in several twin experiments achieved by assimilating two observational datasets, albedo and snow depth, using the PBS scheme on an ensemble with $N_e=1000$ members. Fig. \ref{fig:PBS_abl} shows the posterior annual SMB for the ablation area for the four different scenarios. The prior and posterior CRPS are calculated by comparing the prior and posterior SMB estimates with the synthetic truth over a 12-year period across these scenarios. After assimilation, the average CRPS values (Fig. \ref{fig:CRPS_all}) for the posterior SMB estimates are $0.05$ m for albedo assimilation, $0.03$ for snow depth assimilation, and $0.02$ m for jointly assimilating both observations. For all assimilated observation scenarios this is a marked improvement from the prior CRPS of $0.16$ m. These improvements represent CRPS reductions of $69\%$, $80\%$, and $86\%$, respectively, compared to the prior, showing the enhanced skill of the posterior estimates obtained after data assimilation. On the one hand, snow depth assimilation is particularly effective in bringing the posterior ensemble median SMB closer to the truth. On the other hand, the 95$^\mathrm{th}$ percentile of the posterior ensemble after albedo assimilation often failed to cover the true values. A comparison of error reduction demonstrates (Fig. \ref{fig:PBS_abl}) that joint assimilation of albedo and snow depth provides the most substantial improvements in the performance of posterior SMB simulations. When comparing the four scenarios, it becomes evident that snow depth assimilation performs better under high snowfall conditions, yielding a CRPS of $0.03$ m compared to $0.04$ m in low snowfall conditions. Conversely, albedo assimilation performs better in low snowfall scenarios, with a CRPS of $0.04$ m compared to $0.06$ m under high snowfall conditions.

Fig. \ref{fig:CRPS_all} provides a comprehensive evaluation of the experiments using the PBS assimilation scheme across all scenarios and areas. In both the ablation and equilibrium line altitude areas, the assimilation of either albedo or snow depth substantially reduces CRPS compared to the prior estimates. The average CRPS reduction is $71\%$ and $74\%$, respectively, when albedo and snow depth are assimilated individually. The difference in performance improvement in terms of CRPS between albedo and snow depth assimilation is particularly notable, ranging from a $10\%$ to $-19\%$ difference, especially under the high snowfall scenario. In most experiments, joint assimilation of albedo and snow depth consistently yields the lowest CRPS values. However, in the accumulation area, results indicate an increase in CRPS following snow depth assimilation under low snowfall scenarios, relative to the prior. In contrast, albedo assimilation still improves performance, though the improvements are less pronounced than in the ablation and equilibrium line altitude areas, especially under high snowfall conditions. Similarly, joint assimilation of albedo and snow depth exhibits behavior similar to snow depth assimilation alone in the accumulation area.

The results indicate that the assimilation of joint albedo and snow depth observations within the PBS framework improves the skill of surface mass balance simulations, particularly in scenarios with high snowfall. The results show that snow depth assimilation tends to perform better under high snowfall conditions, while albedo assimilation is more effective under low snowfall scenarios. Moreover, joint assimilation tends to yield the best (including ties) results across the majority of experiments (10 out of 12) in Fig.~\ref{fig:CRPS_all}, providing the greatest improvements both in terms of reducing uncertainty and bringing the posterior closer to the truth. These findings highlight the importance of selecting appropriate observations to assimilate based on specific climatic conditions to optimize the performance of SMB simulations. In particular, the most robust choice is generally joint data assimilation which can automatically handle trade-offs in the information content of different types of observations.

The ES scheme overall exhibits performance similar to that of the PBS scheme after assimilating albedo and snow depth. Joint assimilation yields the best results, while the assimilation of albedo and snow depth individually shows varying outcomes across different scenarios. A detailed comparison of the two assimilation schemes follows in the next section.

\subsection{Comparison of two data assimilation schemes}
\noindent 
We evaluated both the PBS and ES schemes against synthetic truth using an ensemble size of 1000 members for all experiments. Table \ref{tab:pbs_es_crps} presents the improvement in glacier mass balance simulation CRPS performance achieved by the two data assimilation schemes compared to the prior. The values represent the average improvement across four truth scenarios, calculated by comparing the posterior results with the prior. For albedo assimilation, PBS shows a significantly better overall performance compared to ES across all glacier zones. However, for snow depth assimilation, ES performs slightly better than PBS, except in the accumulation zone. Joint assimilation of both albedo and snow depth yields the best performance, regardless of the assimilation method used. In terms of different glacier zones, the performance in the ablation area is generally the best across both data assimilation schemes. In the ELA region, the results slightly underperform those in the ablation area, considering the average performance of three distinct assimilated observation scenarios. The accumulation zone yields the lowest accuracy improvement among the glacier zones for both schemes. Nonetheless, the posterior always improved over the prior in terms of mass balance CRPS. Note that these results represent averages across four scenarios and considerable differences exist between individual scenarios, particularly when assimilating snow depth generated under different snowfall factors.

Fig. \ref{Fig:named_rmse_std_ela_1} presents the posterior annual mass balance results derived from the two data assimilation schemes in the ELA region under a scenario of rapid snow albedo evolution and high snowfall. Joint assimilation under the ES scheme demonstrates the best overall performance, achieving the lowest RMSE and standard deviation compared to other configurations. While the PBS scheme also performs well with joint assimilation, it shows no substantial improvement over the albedo or snow depth assimilation individually, and the variability in standard deviation across years is notably higher than in the ES results. For albedo assimilation, both methods considerably enhance the accuracy and reduce uncertainty compared to the prior. However, the PBS scheme yields slightly lower RMSE and standard deviation than the ES scheme but exhibits higher interannual variability. In contrast, the ES scheme demonstrates a more stable annual performance. Regarding snow depth assimilation, the posterior results from the PBS scheme reveal overconfidence, characterized by an ensemble spread near zero and high annual variability, along with a higher average RMSE than the ES scheme. In comparison, the ES scheme produces a smaller standard deviation and maintains stable annual performance after snow depth assimilation, with no marked interannual fluctuations. However, this advantage may partially stem from the bias introduced by overconfidence, especially influenced by snow depth assimilation. It is also unclear if the higher computational cost of $2N_e$ CryoGrid running with the ES, compared to just $N_e$ with the PBS, justifies the slight gain in performance in this case. 

\begin{table}
   \caption{Comparison of two data assimilation methods in improving the average CRPS of glacier mass balance simulations by assimilating different observations for various glacier zones. The values in the table represent the average CRPS improvement, calculated by comparing the percentage improvement of the posterior CRPS results to that of the prior CRPS results, across all four scenarios.}
    \label{tab:pbs_es_crps} 
    \centering
    \begin{tabular}{l|l c c c}
        \hline
        \textbf{} & \textbf{}  & \textbf{Albedo} & \textbf{Snow depth} & \textbf{Joint} \\ \hline
        \textbf{PBS} & \textbf{ABL}   & 68.4\%             & 77.8\%       & 85.4\%  \\ 
                    & \textbf{ELA}   & 72.2\%             & 66.9\%       & 79.5\%   \\ 
                    & \textbf{ACC}   & 31.0\%             & 19.9\%       & 25.5\%   \\ \hline
        \textbf{ES} & \textbf{ABL}   & 48.3\%             & 79.0\%       & 85.6\%        \\ 
                    & \textbf{ELA}   & 47.9\%             & 67.4\%       & 76.7\%        \\ 
                    & \textbf{ACC}  & 24.1\%             & 11.6\%       & 25.4\%        \\ \hline
    \end{tabular}
    
\end{table}

\begin{figure*}
    \centering
    \hspace*{-0.5 cm}
    \includegraphics[width = 178mm]{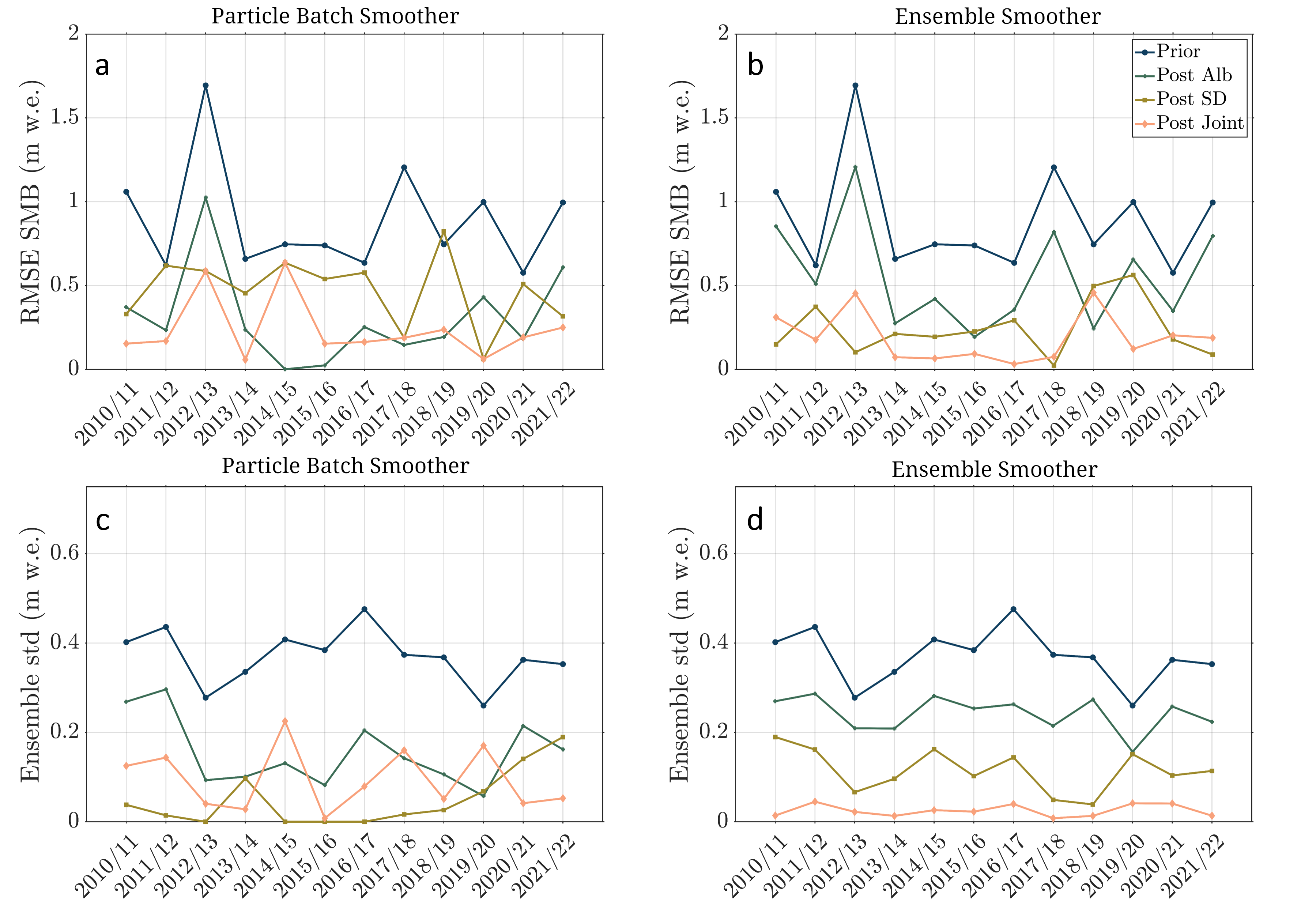}
    \caption{Comparison of the performance of two assimilation schemes applied to the ELA area in terms of RMSE (top row) and ensemble standard deviation (bottom row) for the Particle Batch Smoother (left panels a and c) and the Ensemble Smoother (right panels b and d).}
    \label{Fig:rmse_std_ela_1}
\end{figure*}

\subsection{Sensitivity of data assimilation performance to the ensemble size}
\begin{figure}
    \centering
    \hspace*{-0.5 cm}
    \includegraphics[width = 89mm]{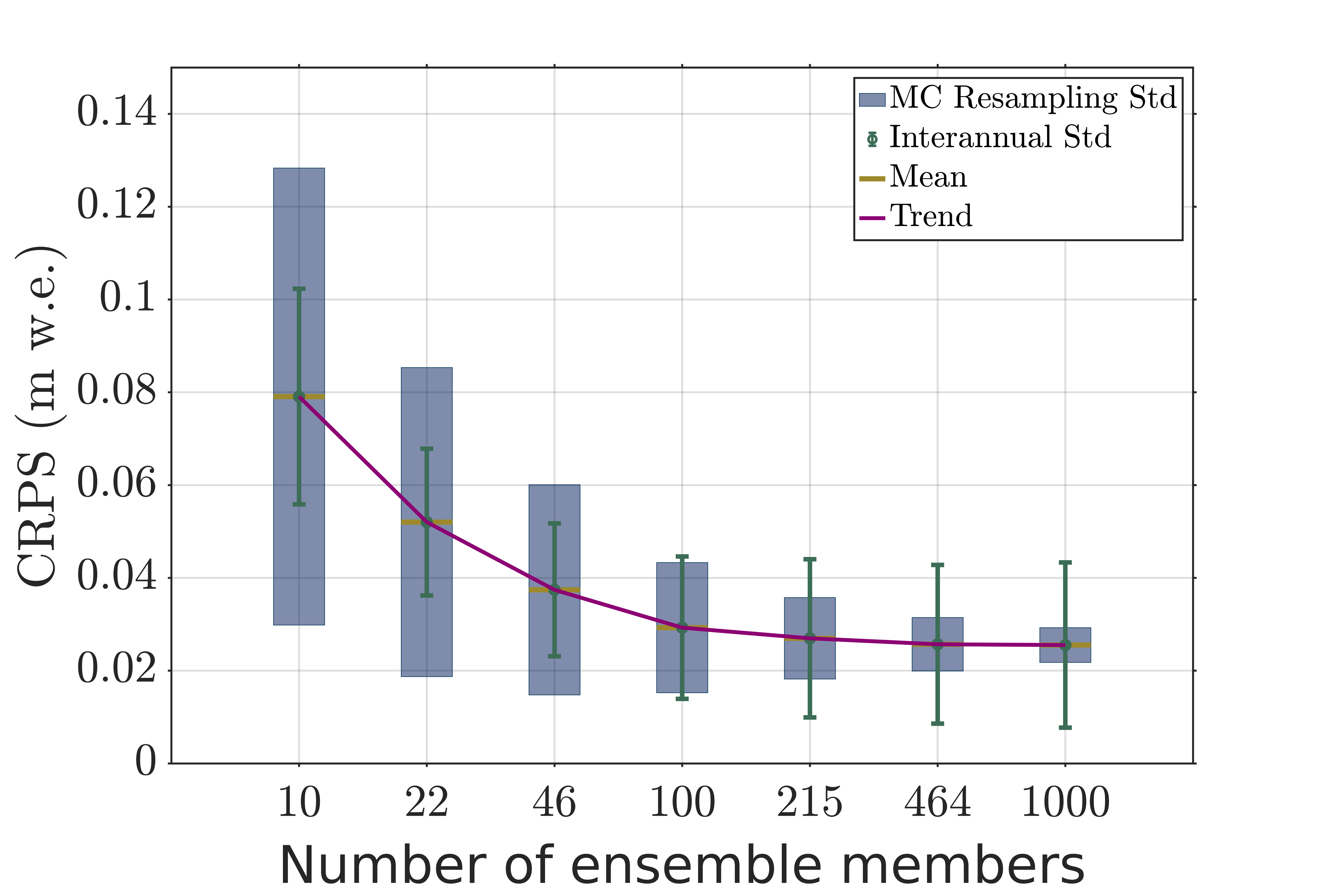}
    \caption{Sensitivity of the posterior surface mass balance CRPS to ensemble size following joint assimilation of albedo and snow depth using the PBS scheme under the scenario of R\&H in the ablation area. For each ensemble size ($N_e$) the CRPS statistics were estimated by resampling with replacement (i.e., bootstrapping) an ensemble of $N_e$ particles from the complete large ensemble ($1000$ members) $100$ times, evaluating the CRPS for each of these $100$ bootstrapped ensembles, and subsequently computing sample statistics.}
    \label{Fig:sensi_abl_jt_crps_1}
\end{figure}

In this section, we present the results of analysing the sensitivity of data assimilation performance to the number of ensemble members $N_e$. The range of ensemble sizes $N_e$ investigated was selected to be regular on a logarithmic scale, generating a vector of seven logarithmically spaced values between $10^1$ and $10^3$ ensemble members. Fig. \ref{Fig:sensi_abl_jt_crps_1} illustrates the mean and variance of CRPS values obtained from $100$ iterations of bootstrapping (resampling with replacement) prior ensembles of variable size $N_e$ from the original large ensemble of $1000$ prior parameters (used in the rest of the study) followed by the assimilation of joint albedo and snow depth under the PBS scheme. The results indicate that, across all experiments, both the average CRPS and its Monte Carlo variance decrease as the ensemble size increases. This shows the classic improvement in performance, both in terms of mean and variance, with increased ensemble size as expected from Monte Carlo methods. Moreover, as expected, the rate of error reduction diminishes considerably, particularly after the ensemble size reaches 100, the mean CRPS starts to show clear convergent behavior towards an asymptote around $0.025$ (m w.e.) with a steadily decreasing variance. However, unlike the Monte Carlo variance, interannual variability remains relatively stable and does not exhibit any clear dependence on ensemble size.

\section{Discussion}

\subsection{Influence of observations on glacier mass balance modeling}
\noindent Across all scenarios and regions, the assimilation of albedo consistently brings the ensemble median of the SMB simulations closer to the true values while effectively reducing the ensemble spread. This improvement is consistent with the findings of \citet{Dumont2012VariationalGlacier}, which demonstrated that assimilating MODIS-derived albedo in a snowpack model improves the accuracy of the SMB simulation for an alpine glacier in the French Alps through variational assimilation. Despite claims to this effect, \citet{Dumont2012VariationalGlacier} did not show how their variational data assimilation scheme constrained uncertainty. In contrast, our ensemble-based data assimilation results show that both the PBS and ES schemes effectively constrain the ensemble, leading to significant reductions in uncertainty. Moreover, unlike variational methods, the ensemble-based schemes pursued herein do not require a differentiable data-generating model and are thus more widely applicable. Fig. \ref{Fig:rmse_std_ela_1} highlights the improvement in accuracy and the reduction in uncertainty achieved by albedo assimilation, with PBS outperforming ES in both metrics. 

\begin{figure*}
    \centering
    \hspace*{-0.5 cm}
    \includegraphics[width = 178mm]{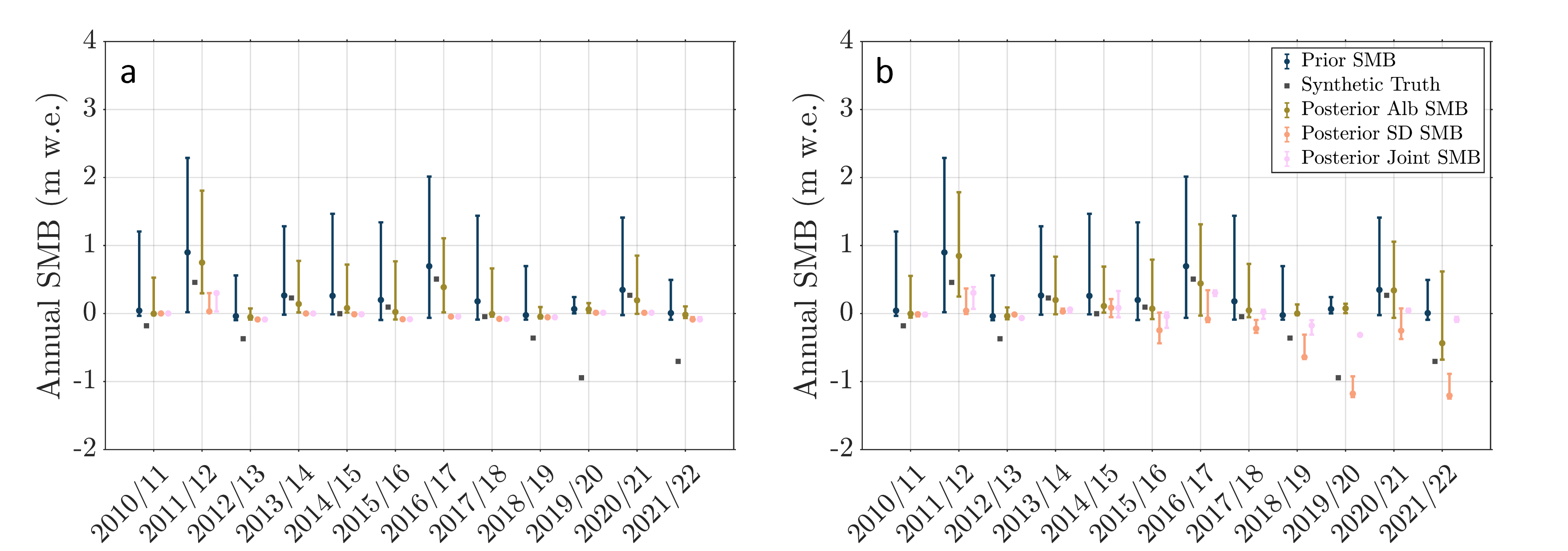}
    \caption{Comparison of the prior, posterior, and true annual surface mass balance in the accumulation area when assimilating different types of observations with the PBS (a) and the ES (b) under a low snowfall factor and slow albedo evolution rate}
    \label{Fig:PBS_ES_acc_2}
\end{figure*}

The impact of snow depth assimilation on SMB simulations exhibits some regional variability, but overall, snow depth assimilation generally enhances SMB accuracy, with more consistent improvements observed outside the accumulation area. Under high snowfall factor scenarios, snow depth assimilation markedly improves SMB simulation accuracy across all regions, aligning with the general findings from  \citet{Landmann2021AssimilatingFilter} for surface mass balance and \citet{Magnusson2017ImprovingFilter} for seasonal snow. 

Our results show a substantial impact of snow depth assimilation on SMB simulations, with an average improvement of 74\% in SMB accuracy in both the ablation and ELA regions. This demonstrates similar performance gains to previous studies. For example, \citet{Landmann2021AssimilatingFilter} reported a relatively low CRPS of $0.012 m w.e.$ mass balance compared with cumulative observations, while \citet{Magnusson2017ImprovingFilter} observed a 64\% reduction in SWE error from snow depth assimilation across 40 sites in Switzerland. Unlike these studies that use particle filtering techniques, we apply the smoothing-based PBS and ES schemes that allow information from the observations to propagate backward in time which has been shown to be advantageous for retrospective snow data assimilation \citep{Alonso-Gonzalez2022TheV1.0}.

Under low snowfall scenarios, snow depth assimilation alone yields less favorable results, particularly in the accumulation area. As illustrated in Fig. \ref{Fig:PBS_ES_acc_2}, posterior estimates in PBS collapse to a single particle with snow depth assimilation in this scenario. This phenomenon is likely due to limitations inherent in the PBS scheme \citep{Robinson2018ImprovingObservations,Pirk2022} and the nature of the low snowfall setting, which produces some SMB truth values that fall outside the prior ensemble range. This discrepancy prevents the posterior from fully encompassing true values, and, when coupled with the ensemble’s overconfidence, results in an increased CRPS due to bias and overconfident predictions. Under the same conditions, ES outperforms PBS due to fundamental differences in both the assumptions and updates steps in these methods \citep{Margulis2015,Aalstad2018Ensemble-basedSites,Alonso-Gonzalez2022TheV1.0}.

To address data availability challenges, we generated synthetic observational data for albedo and snow depth, potentially providing daily coverage over a full year. Subsequently, we applied the specific methods mentioned above to select data points that mimic the temporal availability of ICESat-2 and MODIS measurements. This approach enabled us to control the experimental environment under conditions of parameter uncertainty, thereby facilitating the execution of large ensemble experiments. While this approach theoretically fulfilled continuous data requirements, achieving similar completeness with real observational data remains challenging \citep{Sandven2023SeaSets, Gabarro2023ImprovingGaps}. Satellite-based measurements, such as those from ICESat-2 and MODIS, face limitations due to cloud contamination, which degrades data quality and restricts data acquisition\citep{stby2014SevereSvalbard,Neuenschwander2019CanopyLook, Kotarba2022ImpactMissions}. Additionally, optical satellites that provide albedo data are limited by daylight availability \citep{Wang2018LimitationsRegions}, resulting in data gaps in areas with heavy cloud cover or reduced sunlight. Consequently, while synthetic data can theoretically satisfy continuous data requirements, real-world data collection remains inherently constrained by these observational challenges that we aimed to replicate in the design of our twin experiments.

\subsection{Performance of data assimilation schemes}
For all the given observations and research areas, both data assimilation schemes contribute to considerably reductions in uncertainty and error in SMB simulations. The PBS showed superior performance in albedo assimilation, offering a more confident and accurate ensemble. Conversely, ES generally outperformed PBS in snow depth assimilation scenarios, particularly where the model's prior did not bracket the truth value (Fig. \ref{Fig:PBS_ES_acc_2}). The PBS operates by weighing the ensemble of states based on their likelihood \citep{Margulis2015, Aalstad2018Ensemble-basedSites}, avoiding the need to move particles in parameter space. This results in lower computational demands for state estimation as it only requires one model run per ensemble member. In our study, PBS was particularly effective for albedo assimilation, offering significant uncertainty reduction with less computational effort. However, the performance of PBS can be limited when the true state falls outside the range of the prior ensemble\citep{Robinson2018ImprovingObservations, Pirk2022}, as seen in scenarios with low snowfall where the posterior ensemble sometimes became degenerate and overconfident.

In contrast, ES updates the state by moving particles in parameter space, which can lead to better coverage of the true state, especially when it lies outside the prior ensemble range \citep{vanLeeuwen1996, Evensen2022DataFundamentals}. This adaptability was evident in our results, where ES performed better in assimilating snow depth, particularly under low snowfall scenarios. The ES method requires twice the number of model runs compared to PBS because it requires rerunning the model with updated parameters, which increases computational cost, but can lead to more accurate results in certain scenarios.

Compared to traditional Markov Chain Monte Carlo (MCMC) methods, which involve numerous sequential iterations to converge on a solution, both PBS and ES use parallelizable ensemble approaches that significantly reduce computational time. For instance, in the study of \citet{Rounce2020QuantifyingAsia}, MCMC methods were used to quantify parameter uncertainty in glacier models, involving costly iterations to sample the posterior distribution. Our methods avoid the mainly iterative sampling of MCMC by directly updating an ensemble of parameter vectors, providing a faster convergence to a posterior estimate. While MCMC methods can be very accurate due to their thorough sampling of parameter space, they are often computationally heavy for complex models like CryoGrid, where each model simulation is expensive. Our approach integrates the complexity of CryoGrid with efficient data assimilation methods, allowing for more frequent updates or larger ensembles without a proportional increase in computational demand. This stands in contrast to the recent study of \citet{Sjursen2023BayesianUncertainties}, which employed MCMC for parameter estimation in a simpler mass balance model but nonetheless required careful consideration of computational resources. 

\subsection{Sensitivity to ensemble size}
The sensitivity is evaluated based on two components: resampling (Monte Carlo) variance and interannual variability. As reported in the results, the average CRPS decreases with increasing ensemble size. Notably, the variance of CRPS from interannual variability remains unchanged, whereas the variance of CRPS associated with Monte Carlo resampling error follows the overall decreasing trend of the total error. Interannual variance reflects the natural variability in the system over the different years, capturing the system's response to varying climatic conditions \citep{Malone2019InterannualGlaciers, Wei2019QuantifyingRates}. In this study, all experiments are forced using the same meteorological data source, meaning that interannual variance is inherent to the system and remains unaffected by ensemble size. The problem of small ensemble sizes resulting in large resampling variance is well-documented, as subsets sampled from smaller ensembles may fail to adequately represent the full diversity of a larger ensemble, leading to greater variance in the results \citep{Choi2024SamplingModels}. In our study, as the ensemble size increases to 100, the rate of improvement in CRPS (result accuracy) and the reduction in resampling variance both exhibit a diminishing trend. While larger ensembles generally reduce sampling errors, they come at the cost of increased computational demands \citep{Sacher2008SamplingTheory}. The optimal ensemble size, however, depends on the specific design of the experiment and the acceptable trade-off between computational cost and error tolerance for the user \citep{Milinski2020HowBe}. When the ensemble size reaches 1000, the resampling variation might be expected to approach zero in our experiment design, as the entire large ensemble pool comprises 1000 unique members. However the resampling variation remains non-zero even when the ensemble size is 1000. This is because the CRPS statistics are estimated through a bootstrapping process, where an ensemble of \( N \) particles is resampled \emph{with replacement} from the complete large ensemble. Even when the number of samples matches the original pool size, the randomness introduced by bootstrapping with replacement ensures that the resampled subset does not perfectly replicate the original pool \citep{Davison1997BootstrapApplication}. Some particles may appear multiple times, while others may be excluded entirely. This stochastic nature of the bootstrapping process introduces Monte Carlo sampling error, leading to persistent variability in the results and ensuring a non-zero resampling variance that mimics the actual Monte Carlo variance that would arise when individual particles are sampled multiple times within 1000 ensemble members. The bootstrap technique used herein is a computationally affordable way to probe Monte Carlo sampling error that could otherwise be prohibitively expensive to evaluate in that it would require running multiple distinct large ensembles through CryoGrid.

\section{Conclusions}
In this study, we applied two data assimilation schemes, the PBS and ES, to simulate glacier mass balance across different glacier zones through extensive large ensemble twin experiments.
The posterior results were evaluated by comparing them with synthetic true surface mass balance values using the CRPS metric with the prior CRPS as a reference from which improvement was measured. Cross-comparisons across different scenarios further illustrated the impact of various observational data on mass balance simulations under different assimilation schemes. From this study, the following conclusions can be drawn: 
\begin{itemize}[label=-]
    \item Assimilating albedo generally improves SMB simulation across all glacier zones, with $68.4\%$ improvement by PBS and $48.3\%$ improvement by ES. However, the degree of improvement varies between different glacier areas. In particular, results in the ablation area show an average improvement of $58.4\%$, which is greater than the $27.5\%$ improvement observed in the accumulation area.
    \item The assimilation of snow depth yields results comparable to those of albedo assimilation, particularly in the ablation and ELA zones, $78.4\%$ and $67.1\%$ respectively, averaged by two data assimilation schemes. However, under the low snowfall scenarios within the PBS scheme, methodological limitations cause the posterior results to collapse to a single point in the accumulation zone, resulting in an overly constrained ensemble. This excessive constraint leads to outcomes that are both overconfident and biased.
    \item Both assimilation schemes lead to marked improvements in mass balance simulations. While the PBS outperforms the ES in assimilating albedo, the ES demonstrates superior performance over the PBS when assimilating snow depth.
    \item The joint assimilation of both observation types gives the best performance across all experiments except those given by low snowfall level in the accumulation area. The average improvement in CRPS after joint assimilation across all different glacier areas is $63.5\%$. 
    \item Resampling from the large $1000$ members ensemble using varying ensemble sizes, the rate of improvement, reflected in both the variance of the Monte Carlo resampling and the median CRPS, slows considerably when the ensemble size reaches $100$ indicating diminishing performance gains with further computationally costly increases in ensemble size.
\end{itemize}
%\\
The twin experiments in this study demonstrated strong performance gains across a range of scenarios, including various glacier zones and observational data. This establishes the assimilation approach as effective and suggests that it is potentially transferable for estimating mass balance of all glaciers on Svalbard. However, observational data can be inconsistent in real-world applications, posing further implementation challenges when relying on satellite-based observations due to factors such as gaps and retrieval uncertainty. 

%\\
\section{Data and code availability}
CARRA data was downloaded from the Copernicus Climate Change Service (C3S) Climate Data Store at \\ $https://doi.org/10.24381/cds.d29ad2c6$.
The results contain modified Copernicus Climate Change Service information 2022. Neither the European Commission nor ECMWF is responsible for any use that may be made of the Copernicus information or data it contains. The CryoGrid community model is hosted on Github. The source code is available at \\ $https://github.com/CryoGrid/CryoGridCommunity\_source$. 

\section{Acknowledgements} 
 Wenxue Cao was funded by the European Union’s Horizon 2020 research and innovation program under the Marie Skłodowska-Curie grant agreement number 945371 and the HarSval Bilateral initiative aiming at Harmonisation of the Svalbard cooperation with number UMO-2023/43/7/ST10/00001. Louise S. Schmidt was funded by the Research Council of Norway through the Nansen Legacy project (NFR-276730) and the MAMMAMIA project (NFR-301837). Kristoffer Aalstad acknowledges funding from the ERC-2022-ADG under grant agreement No 01096057 GLACMASS and an ESA CCI Research Fellowship (PATCHES project). The simulations were performed on resources provided by the Department of Geosciences, University of Oslo. Views and opinions expressed are those of the authors only and do not necessarily reflect those of the European Union or the European Research Executive Agency. Neither the European Union nor the granting authority can be held responsible for them.
\section{Author contributions}
Conceptualization was by WC, KA, LSS, and TVS. Data curation was by WC and LSS. Formal analysis was by WC, KA, and TVS. Funding acquisition was by TVS. Methodology was by WC, KA, LSS, and SW. Supervision was by TVS, KA and LSS. Visualization was by WC and KA. Writing- original draft was by WC with key contributions from KA and edited by all co-authors. 

\bibliography{Cao2025arxiv}
\bibliographystyle{igs}  

\end{document}